\documentstyle[epsfig,sprocl]{article}
\begin{document}
\begin{center}
{\Large Structural Studies of Many-Body Systems and
$(e,e^{\prime}p)$
Reaction Cross Sections}\\
\vspace{.5cm} A.N. Antonov$^{1,2}$, M.K. Gaidarov$^{1}$, M.V.
Ivanov$^{1}$, K.A. Pavlova$^{1}$\\
\vspace{.2cm}
{$^{1}$ \it Institute of Nuclear Research and Nuclear Energy,\\
          Bulgarian Academy of Sciences, Sofia 1784, Bulgaria}\\
{$^{2}$ \it Departamento de Fisica Atomica, Molecular y Nuclear,\\
Facultad de Ciencias  Fisicas, Universidad Complutense de
Madrid,\\
Madrid E-28040, Spain}\\
\vspace{.4cm}
C. Giusti\\
\vspace{.2cm} {\it Dipartimento di Fisica Nucleare e Teorica,
Universit\`a di Pavia,\\
Istituto Nazionale di Fisica Nucleare, Sezione di Pavia, Pavia,
Italy}
\end{center}
\vspace{.5cm}

\begin{abstract}
Studies of one-body density matrices (ODM) are performed in
various correlation methods, such as the Jastrow method, the
correlated basis function method, the Green's function method and
the generator coordinate method aiming to extract the absolute
spectroscopic factors and overlap functions (OF) for one-nucleon
removal reactions from the ODM of the target nucleus. The
advantage of this method is that it avoids the complicated task of
calculating the total nuclear spectral function. The procedure for
extracting bound-state OF's has been applied to make calculations
of the cross sections of the $(e,e^{\prime}p)$ reaction on the
closed-shell nuclei $^{16}$O and $^{40}$Ca as well as on the
open-shell nucleus $^{32}$S consistently (using the same OF's)
with the cross sections of $(p,d)$ and $(\gamma,p)$ reactions on
the same nuclei. The analyses of the reaction cross sections and
the spectroscopic factors and the comparison with the experimental
data show the particular importance of these OF's, since they
contain effects of nucleon correlations (short-range and/or
long-range) which are accounted for to different extent in the
theoretical methods considered.
\end{abstract}

\section{Introduction}
The strong short-range and tensor components of the
nucleon-nucleon (NN) interactions induce correlations in the
nuclear wave function which are going beyond the
independent-particle approximation. Therefore, it has always been
a point of experimental and theoretical interest to find
observables which reflect these correlations in a unambiguous way.
In this sense both, the overlap functions and single-nucleon
spectroscopic factors, have attracted much attention in analyzing
the empirical data from one-nucleon removal reactions, such as
$(e,e^{\prime }p)$, $(p,d)$, $(d,^{3}He)$, and also in other
domains of many-body physics, as e.g. atomic and molecular physics
\cite{An93}.

Recently, a general procedure has been adopted \cite{Vn93} to
extract the bound-state overlap functions and the associated
spectroscopic factors and separation energies on the base of the
ground-state (g.s.) one-body density matrix. This makes it
possible to investigate the effects of the various types of NN
correlations included in the ODM on these structural quantities.
Of course, the general success of the above procedure depends
strongly on the availability of realistic ODM's.

Initially, the method for extracting bound-state overlap functions
has been applied in \cite{Sto96} to a model ODM \cite{Sto93}
accounting for the short-range nucleon correlations within the
Jastrow correlation method. The resulting OF's have been used
\cite{Di97} to study one-nucleon removal processes in contrast to
the mean-field approaches which account for the nucleon
correlations by modifying the mean-field potentials. The results
obtained for the differential cross sections of $^{16}O(p,d)$ and
$^{40}Ca(p,d)$ pick-up reactions at various incident energies
demonstrated that the OF's can be applied as realistic form
factors to evaluate absolute cross sections of such reactions. The
analysis of single-particle (s.p.) OF's has been extended to more
realistic ODM's emerging from the correlated basis function (CBF)
method \cite{Sa96,Sa97}, the Green function method (GFM)
\cite{Po96} and the generator coordinate method (GCM)
\cite{An93,An88}. In addition, ODM's of open-shell nuclei deduced
from Jastrow-type calculations have been used \cite{Mous2000}. We
have chosen the CBF theory since it is particularly suitable for
the study of the short-range correlations (SRC) in nuclei. The CBF
calculations have recently been extended to medium-heavy
doubly-closed shell nuclei \cite{Sa96,Sa97} using various levels
of the Fermi hypernetted chain approximation \cite{Sa96}. The GFM
\cite{Po96,Di92} provides detailed information on the spectral
functions and nucleon momentum distributions predicting the
largest effects of the short-range and tensor correlations at high
momentum and energy. The results on the one- and two-body density
and momentum distributions, occupation probabilities and natural
orbitals obtained within the GCM using various construction
potentials \cite{Iva2000} have shown that the NN correlations
accounted for in this method are different from the short-range
ones and are rather related to the collective motion of the
nucleons.

The main aim of the present work is to study the effects of the NN
correlations included in the methods mentioned above on the
behavior of the bound-state proton and neutron overlap functions
in closed- as well as open-shell nuclei and of the related
one-nucleon removal reaction cross sections. Such an investigation
allows to examine the relationship between the ODM and the
associated overlap functions within the correlation methods used
and also to clarify the importance of the effects of NN
correlations on the overlap functions and the reaction cross
sections.

\section{Overlap functions and their relationship
with the one-body density matrix}

For a correct calculation of the cross section of nuclear
reactions with one-neutron or one-proton removal from the target
nucleus, the corresponding OF's for the neutron and proton bound
states must be used in the reaction amplitudes. Here we would like
to remind that the single-particle OF's are defined by the overlap
integrals between the eigenstates of the $A$-particle and the
$(A-1)$-particle systems:
\begin{equation}
\phi _{\alpha }({\bf r})=\langle \Psi _{\alpha }^{(A-1)}|a({\bf
r})|\Psi ^{(A)}\rangle , \label{eq:OF}
\end{equation}
where $a({\bf r})$ is the annihilation operator for a nucleon with
spatial coordinate ${\bf r}$ (spin and isospin operators are
implied). In the mean-field approximation $\Psi ^{(A)}$ and $\Psi
_{\alpha }^{(A-1)}$ are single Slater determinants, and the
overlap functions are identical with the mean-field s.p. wave
functions, while in the presence of correlations both $\Psi
^{(A)}$ and $\Psi _{\alpha }^{(A-1)}$ are complicated
superpositions of Slater determinants. In general, the overlap
functions $(\ref{eq:OF})$ are not orthogonal. Their norm defines
the spectroscopic factor
\begin{equation}
S_{\alpha }=\langle \phi _{\alpha }|\phi _{\alpha }\rangle .
\label{eq:spf}
\end{equation}
The normalized to unity OF associated with the state $\alpha $
then reads
\begin{equation}
\tilde{\phi}_{\alpha }({\bf r})=S_{\alpha }^{-1/2}\phi _{\alpha
}({\bf r}). \label{eq:NOF}
\end{equation}
The ODM can be expressed in terms of the OF's in the form:
\begin{equation}
\rho ({\bf r},{\bf r^{\prime }})=\sum_{\alpha }\phi _{\alpha }^{*}({\bf r}%
)\phi _{\alpha }({\bf r^{\prime }})=\sum_{\alpha }S_{\alpha }\tilde{\phi}%
_{\alpha }^{*}({\bf r})\tilde{\phi}_{\alpha }({\bf r^{\prime }}).
\label{eq:rho}
\end{equation}
The asymptotic behavior of the radial part of the neutron OF for
the bound states of the $(A-1)$-system is given by \cite{Vn93}:
\begin{equation}
\phi_{nlj}(r)\rightarrow C_{nlj}\exp(-k_{nlj}r)/r,
\label{eq:nasym}
\end{equation}
where $k_{nlj}$ is related to the neutron separation energy
\begin{equation}
k_{nlj}=\frac {\sqrt{2m\epsilon_{nlj}}}{\hbar}, \;\;\;
\epsilon_{nlj}=E_{nlj}^{(A-1)}-E_{0}^{A}. \label{eq:decay}
\end{equation}
For proton bound states, due to an additional long-range part of
the interaction originating from the Coulomb one, the asymptotic
behavior of the radial part of the corresponding proton OF's reads
\begin{equation}
\phi_{nlj}(r)\rightarrow C_{nlj}\exp[-k_{nlj} r-\eta \ln
(2k_{nlj}r)]/r, \label{eq:pasym}
\end{equation}
where $\eta $ is the Coulomb (or Sommerfeld) parameter and
$k_{nlj}$ in (\ref{eq:decay}) contains in this case the mass of
the proton and the proton separation energy.

Taking into account Eqs. (\ref{eq:rho}) and (\ref{eq:nasym}), the
lowest ($n=n_{0}$) neutron bound-state $lj$-overlap function is
determined by the asymptotic behavior of the associated partial
contribution of the radial ODM $\rho_{lj}(r,r^{\prime})$
($r^{\prime}=a\rightarrow \infty $) as
\begin{equation}
\phi _{n_{0}lj}(r)={\frac{{\rho _{lj}(r,a)}}{{C_{n_{0}lj}~\exp
(-k_{n_{0}lj}\,a})/a}}~, \label{eq:nof}
\end{equation}
where the constants ${C_{n_{0}lj}}$ and ${k_{n_{0}lj}}$ are
completely determined by $\rho_{lj}(a,a)$. In this way the
separation energy $\epsilon_{n_{0}lj}$ and the spectroscopic
factor $S_{n_{0}lj}$ can be determined as well. Similar expression
for the lowest proton bound-state OF can be obtained having in
mind its proper asymptotic behavior (\ref{eq:pasym}).

\section{Results for the cross sections of $(e,e^{\prime}p)$ reactions
on $^{16}$O, $^{40}$Ca and $^{32}$S nuclei}

The inclusion of short-range as well as tensor correlations leads
to an enhancement of the values of the overlap functions in the
interior region and a depletion in the tail region in the
coordinate space \cite{Gai99}. In the momentum space this leads to
a slight redistribution of the strength from the low- to the
high-momentum region. The calculated  spectroscopic factors (SF),
however, differ significantly from the mean--field value. The
depletion in the tail region of the OF's which is due to the NN
correlations leads to lower values of the SF's in comparison with
their mean--filed values. The SF's deduced from the calculations
with different ODM are listed in Table 1. Only short--range
central correlations are included in the ODM in
\cite{Sto96,Sa96,Chr86}, whereas also tensor correlations are
taken into account in \cite{Po96,Vn97}. It was found that
correlation effects on the spectroscopic factor of the hole states
are dominated by the tensor channel of the interaction
\cite{Gai99}.
\vspace{.5cm}\\
\noindent {\bf Table 1:} Spectroscopic factors for the $p_{1/2}$
and $p_{3/2}$ quasihole states in $^{16}O$: column I--deduced from
the calculations with different ODM of $^{16}$O; column
II--additional reduction factors determined through a comparison
between the ($e,e'p$) data of \cite{Leuschner} and the reduced
cross sections calculated in DWIA with the different overlap
functions; column III--total spectroscopic factors obtained from
the product of the factors in I and II.

\begin{center}
\begin{tabular}{cccccccc}
\hline\hline
&  \multicolumn{3}{c}{$1p_{1/2}$} & &  \multicolumn{3}{c}{$1p_{3/2}$} \\
\cline{2-4} \cline{6-8}
ODM & I   & II  & III & & I   & II  & III \\
\hline
HF                       & 1.000 & 0.750 & 0.750 & &  1.000 & 0.550 & 0.550\\
JCM \cite{Sto96}         & 0.953 & 0.825 & 0.786 & &  0.953 & 0.600 & 0.572\\
CBF \cite{Sa96}          & 0.981 & 0.900 & 0.883 & &  0.981 & 0.600 & 0.589\\
CBF \cite{Sa97}          & 0.983 & 0.880 & 0.865 & &  0.983 & 0.630 & 0.619\\
CBF \cite{Vn97}          & 0.912 & 0.850 & 0.775 & &  0.909 & 0.780 & 0.709\\
GFM \cite{Po96}          & 0.905 & 0.800 & 0.724 & &  0.915 & 0.625 & 0.572\\
GCM \cite{Chr86}         & 0.988 & 0.700 & 0.692 & &  0.988 & 0.500 & 0.494\\
\hline\hline
\end{tabular}
\end{center}
\vspace{.5cm}

The reduced cross sections of $(e,e^{\prime}p)$ knockout reactions
have been calculated with the code DWEEPY \cite{DWEEPY}, which is
based on the nonrelativistic distorted wave impulse approximation
(DWIA) description of the nucleon knockout process and includes
final-state interactions and Coulomb distortion of the electron
waves \cite{Oxford}. The latter has been treated with a
high-energy expansion in inverse powers of the electron energy
\cite{DWEEPY}. In the standard DWIA approach, however,
phenomenological s.p. wave functions were used, with some
parameters fitted to the data. In this paper the results have been
obtained with theoretically calculated overlap functions which do
not include free parameters.

\begin{figure}[htb]
\centerline{\epsfig{file=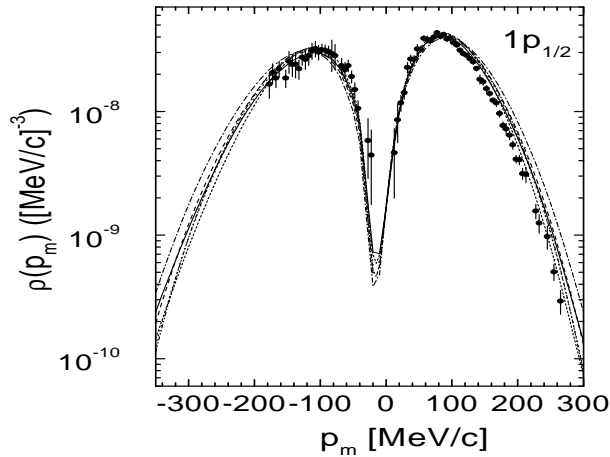,width=80mm,height=6.0cm}}
\caption[]{Reduced cross section of the $^{16}$O($e,e'p$) reaction
as a function of the missing momentum $p_{\mathrm m}$ for the
transition to the $1/2^{-}$ ground state of $^{15}$N in parallel
kinematics, with $E_0=520.6$ MeV and an outgoing proton energy of
90 MeV. The optical potential is from \cite{Schwandt} (see Table
III of \cite{Leuschner}). Overlap functions are derived from the
ODM of GFM \cite{Po96} (solid line), CBF \cite{Vn97} (long-dashed
line), CBF \cite{Sa96} (dot-dashed line), JCM \cite{Sto96} (double
dot-dashed line) and GCM \cite{Chr86} (short-dashed line). The
dotted line is calculated with the HF wave function. The
experimental data are taken from \cite{Leuschner}. The theoretical
results have been multiplied by the reduction factor given in
column II of Table I.} \label{eepo}
\end{figure}

The reduced cross sections $\rho (p_{\mathrm m})$ for the
$^{16}$O($e,e'p$) reaction as a function of the missing momentum
$p_{\mathrm m}$ and for the transitions to the $1/2^{-}$ ground
state is displayed in Figure \ref{eepo}. As can be seen the cross
sections are sensitive to the shape of the various overlap
functions used. The differences  are considerable at large values
of $p_{\mathrm m}$, where the cross section is several orders of
magnitude lower than in the maximum region. The deviations of the
various results at large values of $p_{\mathrm m}$ are related to
different accounting for the short-range NN correlations within
the correlation methods used. SRC are particularly important in
one-nucleon emission at large missing momenta and energy
\cite{PR,MD}. At high missing energies, however, other competing
processes are also present and a clear identification of SRC can
better be made by means of two-nucleon knockout reactions. At low
missing-energy values measurements over an extended range of
missing momenta, in particular at large values, where the SRC
effects seem to be more sizable, can test the various s.p. overlap
functions and NN correlations.

In order to reproduce the size of the experimental cross section a
reduction factor has been applied to the theoretical results in
Fig. \ref{eepo}. These factors, which have been obtained by a fit
of the calculated reduced cross sections to the data over the
whole missing-momentum range considered in the experiment, are
also listed in Table I (column II). In general, a fair agreement
with the shape of the experimental distribution is achieved. The
results, however, are also sensitive to details of the various
overlap functions. The best agreement with the data, for both
transitions, is obtained with the overlap functions \cite{Gai99}
emerging from the ODM calculated within the GFM \cite{Po96}. This
is due to the substantial realistic inclusion of short-range as
well as tensor correlations in the ODM. The calculations based on
the Green function theory \cite{Gai99} have shown that about 10
$\%$ of the $1p$ strength is removed by these correlations. An
excellent agreement with the data is obtained also by using the
overlap function from the GCM \cite{Chr86}, which does not include
tensor correlations. The reduced cross section calculated on the
base of the OF from \cite{Vn97} is in accordance with the data for
the $1/2^{-}$ state. The shape of the experimental reduced cross
sections can adequately be described also by the HF wave
functions, in particular for $p_{\mathrm m}\leq 150$ MeV/$c$.

The fact that in general our results overestimate the data (and a
reduction factor has to be applied to the calculated cross
sections to reproduce them correctly) may be explained having in
mind that our overlap functions are deduced from calculations
including only SRC but not long--range correlations (LRC). The
reduction factor can thus be considered as a further spectroscopic
factor reflecting the depletion of the quasihole state produced by
LRC. Of course, the discrepancy with the data can be due also to
other effects not included or not adequately described by the
theoretical treatment. We note, however, that the reduction
factors applied here to the calculated cross sections are not the
result of a precise theoretical calculation. They have been
obtained by a fit to the data and have only an indicative meaning.
Small variations within 10-15\% around their values would not
significantly change the comparison with data. In any case the
reduction factors should mostly be ascribed to LRC, but for the HF
wave function, which does not contain any kind of correlations.
For this wave function the reduction factor accounts for both LRC
and SRC. It is interesting to note that in the calculations with
the correlated overlap functions the reduction factors for the
$1p_{1/2}$ state turn out to be close to the spectroscopic factor
(0.83) obtained in the theoretical approach of \cite{Amir} where
only LRC are included.

In Table I we give, in addition, in column III the factor obtained
by the product of the two factors in columns I and II. This factor
can be considered as a total spectroscopic factor and can be
attributed to the combined effect of SRC and LRC. Indeed for
$1p_{1/2}$ these factors are in reasonable agreement with the
spectroscopic factor value (0.76) calculated in \cite{Geurts},
where both SRC and LRC are included.

\begin{figure}[htb]
\center{\includegraphics[height=40mm]{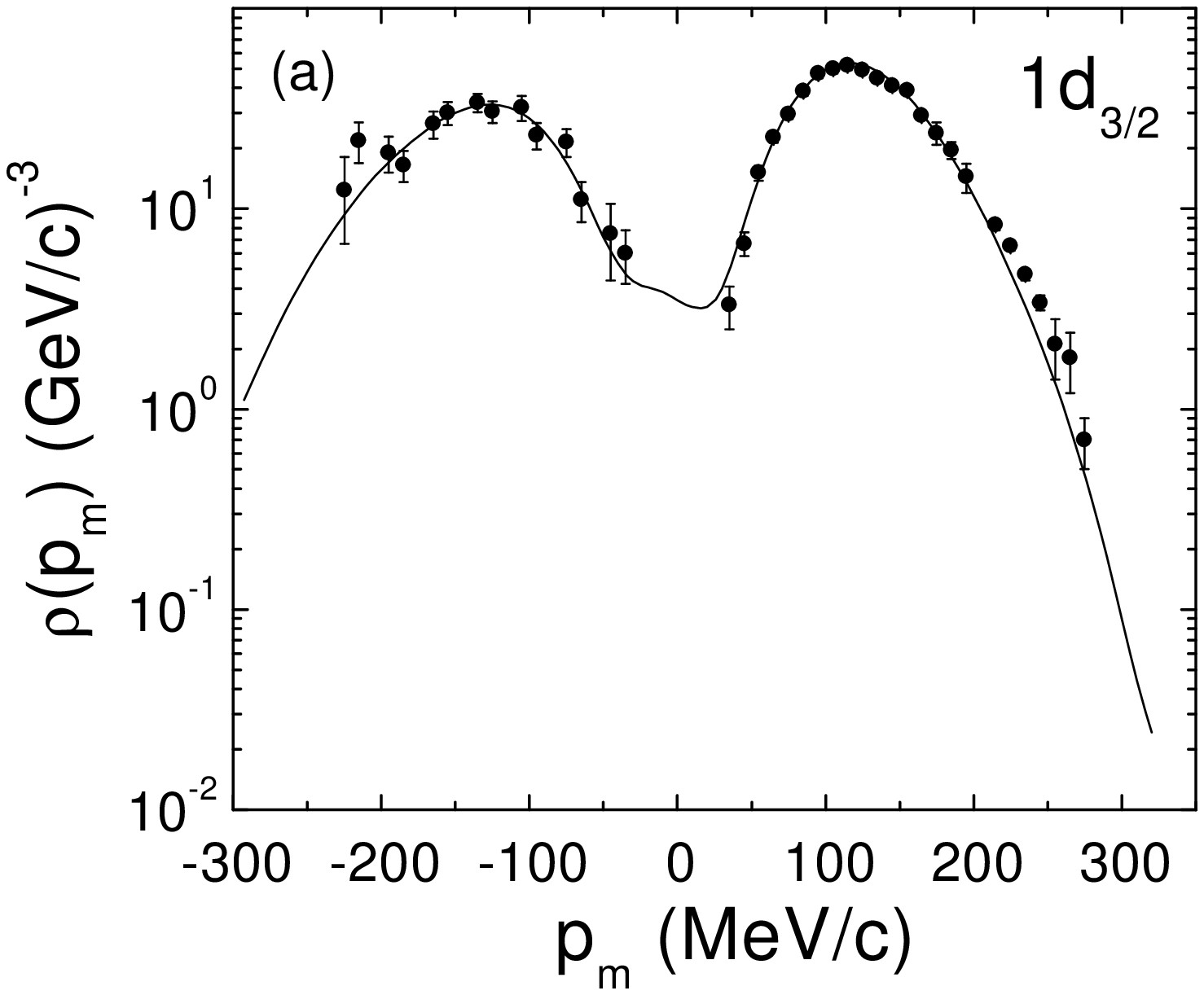} \hspace{10mm}
\includegraphics[height=40mm]{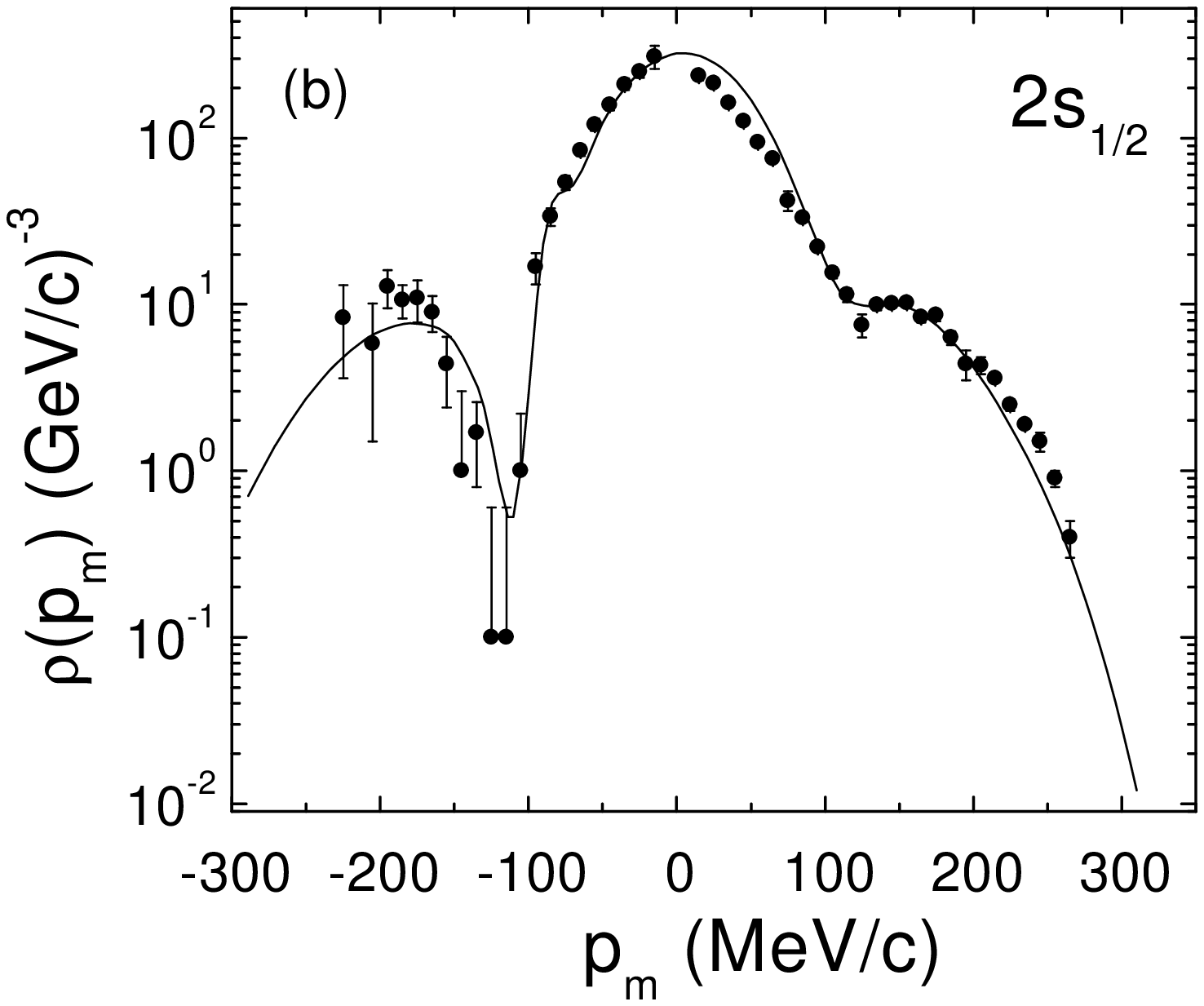}}
\caption[]{Reduced cross section of the $^{40}$Ca($e,e'p$)
reaction for the transitions to the $3/2^{+}$ ground (a) and to
the first $1/2^{+}$ excited (b) states of $^{39}$K in parallel
kinematics. Overlap functions is derived from the ODM of GCM
\cite{Chr86,Iva2001} (solid lines). The experimental data are
taken from \cite{Kra90}.
\label{eepca}}
\end{figure}

\begin{figure}[htb]
\centerline{\epsfig{file=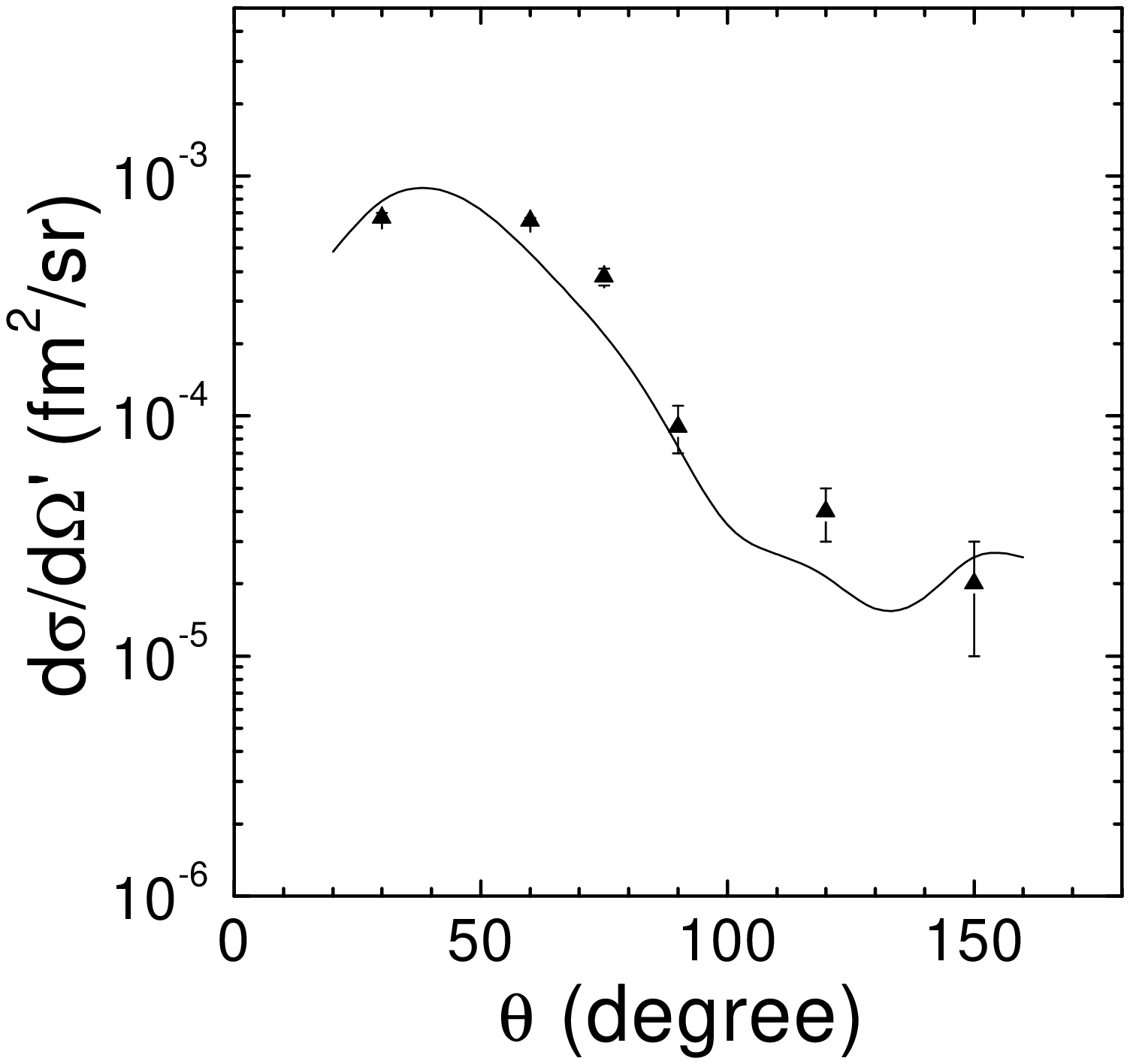,width=60mm,height=5.0cm}}
\caption[]{Angular distribution of the $^{40}$Ca($\gamma,p$)
reaction for the transition to the $3/2^{+}$ ground state of
$^{39}$K at $E_\gamma = 60$ MeV. Overlap functions is derived from
the ODM of GCM \cite{Chr86,Iva2001} (solid line). The experimental
data are taken from \cite{Abe92}.
\label{gpca}}
\end{figure}

A consistent analysis of the cross sections of $(e,e'p)$ and
$(\gamma,p)$ on $^{40}$Ca by means of the same overlap functions
is given in Figures \ref{eepca} and \ref{gpca}, respectively. The
calculated cross sections (presented by solid lines in Fig.
\ref{eepca} and Fig. \ref{gpca}) are generally in good agreement
with the experimental data \cite{Kra90,Abe92}. The important role
of the additional reduction spectroscopic factors applied to the
present calculations, namely 0.55 for the $3/2^{+}$ ground state
and 0.50 for the $1/2^{+}$ excited state, is pointed out. Apart
from the short-range and tensor correlations studied in previous
papers \cite{Gai99,Gai2000}, in this work we looked into the role
of correlations caused by the collective nucleon motion. The
results indicate that the effects of NN correlations taken into
account within our approach and which are of long-range type are
of significant importance for the correct analysis of the
processes considered.

\begin{figure}[htb]
\centerline{\epsfig{file=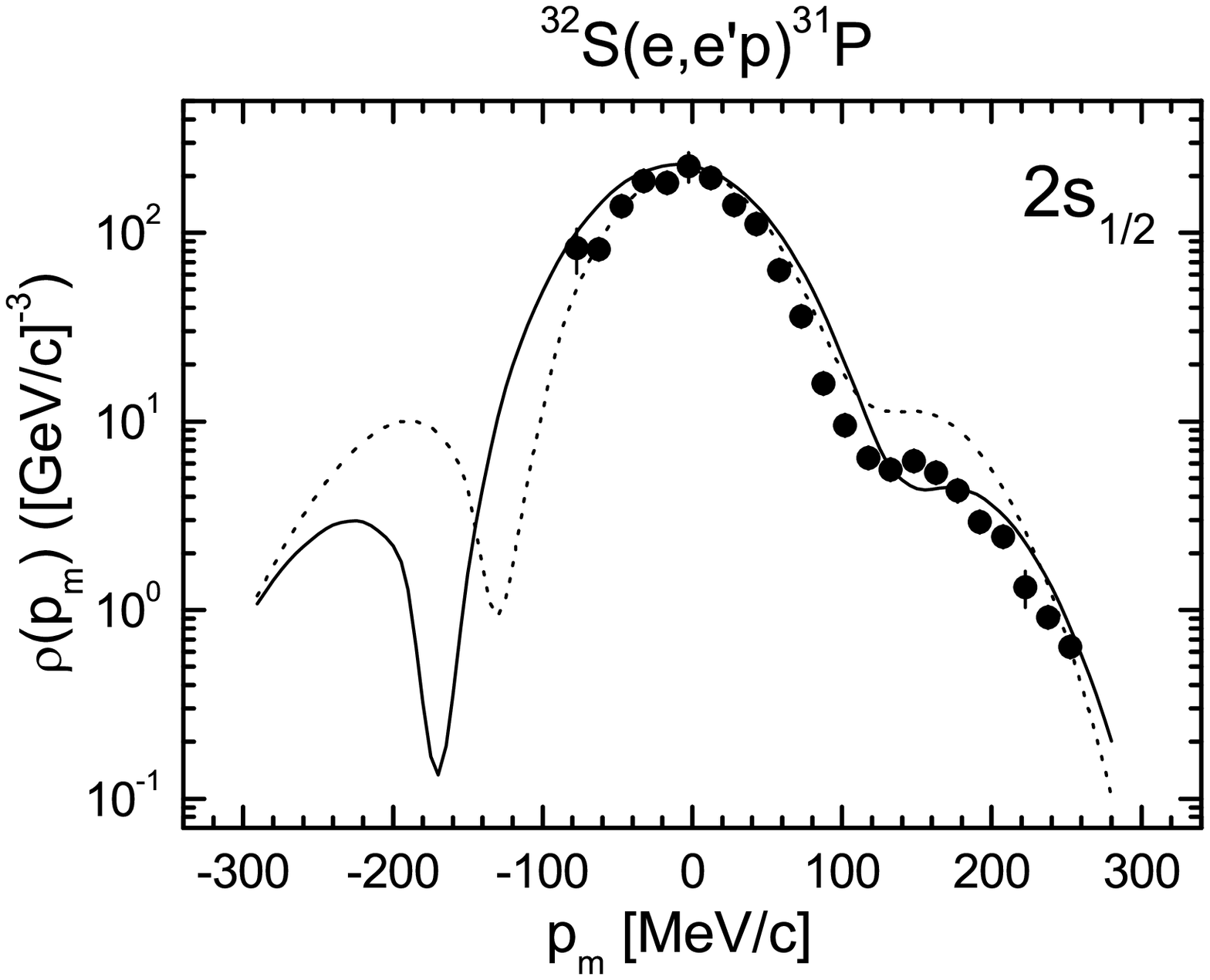,width=80mm,height=6cm}}
\caption[]{Reduced cross section of the $^{32}$S($e,e'p$) reaction
as a function of the missing momentum $p_{\mathrm m}$ for the
transition to the $1/2^{+}$ ground state of $^{31}$P. The proton
overlap function is derived from the ODM (solid line). The result
with the uncorrelated (HO) wave function is given by dotted line.
The experimental data (full circles) are taken from \cite{Wes92}.}
\label{eeps}
\end{figure}

An example of electron induced proton knockout from $^{32}$S for
the transition to the ground $2s_{1/2}$ state of $^{31}$P is
illustrated in Figure \ref{eeps}. In the figure the result
obtained with the proton OF for the $2s$ state of $^{32}$S and the
optical potential from \cite{Schwandt} is compared with the NIKHEF
data from \cite{Wes92}. A reasonable agreement with the
experimental data for the reduced cross section is obtained. We
emphasize that in the present work the OF theoretically calculated
on the basis of the Jastrow-type ODM of $^{32}$S does not contain
free parameters. It can be seen from Fig. \ref{eeps} that our
spectroscopic factor of 0.5648 gives a good agreement with the
size of the experimental cross section and, in addition, it is in
accordance with the integrated strength for the valence $2s_{1/2}$
shell in $^{32}$S \cite{Wes92} which amounts to 65(7)\% of the SP
strength obtained using the shell-model bound-state function. The
result for the $^{32}$S$(e,e^{\prime}p)$ cross section obtained
with the harmonic-oscillator bound-state wave function is also
illustrated in Fig. \ref{eeps}. It has been computed with the same
oscillator parameter value $b$=2 fm for the $2s_{1/2}$
ground-state wave function as in the original calculations of the
ODM without SRC \cite{Mous2000}. In order to perform a consistent
comparison with the result when considering theoretically
calculated overlap function, we have applied the same
spectroscopic factor of 0.5648. In this case, the size of the
reduced cross section is also reproduced, but the HO wave function
gives much worse description of the experimental data. Thus, the
comparison made in Fig. \ref{eeps} shows the important role of the
SRC accounted for in our approach for the correct description of
knockout reactions.

\section{Conclusions}
The s.p. overlap functions calculated on the basis of ODM for the
ground state of closed- and open-shell nuclei emerging from
different correlation methods have been used to calculate the
cross sections of the ($e,e'p$), ($\gamma,p$) and $(p,d)$
reactions. The theoretical results for the cross sections show
that they are sensitive to the shape of the different OF's and are
generally able to reproduce the shape of the experimental cross
sections. In order to describe correctly the experimental data on
the reduced cross sections a reduction factor must be applied to
the calculated cross sections. The fact that it is consistent in
different nucleon removal reactions gives a more profound
theoretical meaning to this parameter. The results indicate that
the effects of SRC correlations taken into account within CBF and
GFM and of correlations accounted for in GCM which are of
long-range type are of significant importance for the correct
analysis of the processes considered.

The authors thank the Bulgarian National Science Foundation which
partly supported this work under the Contract No.$\Phi $--905. One
of the authors (A.N.A.) is grateful for support during his stay at
the Complutense University of Madrid to the State Secretariat of
Education and Universities of Spain (Nº.Ref.SAB2001-0030).


\begin{thebibliography}{99}
\bibitem{An93}  A.N. Antonov, P.E. Hodgson, and I.Zh. Petkov, {\it Nucleon
Correlations in Nuclei} (Springer-Verlag, Berlin, 1993).

\bibitem{Vn93}  D. Van Neck, M. Waroquier, and K. Heyde, {\it Phys. Lett.
B} {\bf 314} (1993) 255.

\bibitem{An95}  A.N. Antonov, M.V. Stoitsov, M.K. Gaidarov, S.S. Dimitrova,
and P.E. Hodgson, {\it J. Phys. G} {\bf 21} (1995) 1333.

\bibitem{Sto96}  M.V. Stoitsov, S.S. Dimitrova, and A.N. Antonov, {\it Phys. Rev.
C} {\bf 53} (1996) 1254.

\bibitem{Sto93}  M.V. Stoitsov, A.N. Antonov, and S.S. Dimitrova, {\it Phys. Rev.
C} {\bf 47} (1993) R455; {\it Phys. Rev. C} {\bf 48} (1993) 74;
{\it Z. Phys. A} {\bf 345} (1993) 359.

\bibitem{Di97}  S.S. Dimitrova, M.K. Gaidarov, A.N. Antonov, M.V. Stoitsov,
P.E. Hodgson, V.K. Lukyanov, E.V. Zemlyanaya, and G.Z. Krumova,
{\it J. Phys. G} {\bf 23} (1997) 1685.

\bibitem{Sa96}  F. Arias de Saavedra, G. Co', A. Fabrocini, and S. Fantoni,
{\it Nucl. Phys. A} {\bf 605} (1996) 359.

\bibitem{Sa97}  F. Arias de Saavedra, G. Co', and M.M. Renis, {\it Phys. Rev.
C} {\bf 55} (1997) 673.

\bibitem{Vn97} D. Van Neck, L. Van Daele, Y. Dewulf, and M. Waroquier,
{\it Phys. Rev. C} {\bf 56} (1997) 1398.

\bibitem{Po96}  A. Polls, H. M\"{u}ther and W.H. Dickhoff, {\it Proceedings
of Conference on Perspectives in Nuclear Physics at Intermediate
Energies}, Trieste, 1995, edited by S. Boffi, C. Ciofi degli Atti,
and M.M. Giannini, p.308 (1996, World Scientific).

\bibitem{An88}  A.N. Antonov, P.E. Hodgson, and I.Zh. Petkov, {\it Nucleon
Momentum and Density Distributions} (Clarendon Press, Oxford,
1988).

\bibitem{Mous2000} Ch.C. Moustakidis and S.E. Massen, {\it Phys. Rev. C} {\bf 62}
(2000) 034318.

\bibitem{Di92}  W.H. Dickhoff and H. M\"{u}ther, {\it Rep. Prog. Phys.} {\bf 55}
(1992) 1947.

\bibitem{Iva2000} M.V. Ivanov, A.N. Antonov, and M.K. Gaidarov,
{\it Int. J. Mod. Phys. E} {\bf 9} (2000) 339.

\bibitem{Gai99} M.K. Gaidarov, K.A. Pavlova, S.S. Dimitrova, M.V. Stoitsov,
A.N. Antonov, D. Van Neck, and H. M\"{u}ther, {\it Phys. Rev. C}
{\bf 60} (1999) 024312.

\bibitem{Chr86} A.N. Antonov, Chr.V. Christov, and I.Zh. Petkov,
{\it Nuovo Cim. A} {\bf 91} (1986) 119; A.N. Antonov, I.S. Bonev,
Chr.V. Christov, and I.Zh. Petkov, {\it Nuovo Cim. A} {\bf 100}
(1988) 779; A.N. Antonov, I.S. Bonev, and I.Zh. Petkov, {\it Bulg.
J. Phys.} {\bf 18} (1991) 169; A.N. Antonov, I.S. Bonev, Chr.V.
Christov, and I.Zh. Petkov, {\it Nuovo Cim. A} {\bf 103} (1990)
1287.

\bibitem{Leuschner} M. Leuschner, J.R. Calarco, F.W. Hersman, E. Jans, G.J. Kramer, L.
Lapik\'{a}s, G. van der Steenhoven, P.K.A. de Witt Huberts, H.P.
Blok, N. Kalantar-Nayestanaki, and J. Friedrich, {\it Phys. Rev.
C} {\bf 49} (1994) 955.

\bibitem{DWEEPY} C. Giusti and F.D. Pacati, {\it Nucl. Phys. A} {\bf 473} (1987) 717;
{\it Nucl. Phys. A} {\bf 485} (1988) 461.

\bibitem{Oxford} S. Boffi, C. Giusti, F.D. Pacati, and M. Radici,
{\it Electromagnetic Response of Atomic Nuclei, Oxford Studies in
Nuclear Physics} (Clarendon Press, Oxford, 1996).

\bibitem{PR} A. Polls, M. Radici, S. Boffi, W.H. Dickhoff, and H. M\"{uther},
{\it Phys. Rev. C} {\bf 55} (1997) 810.

\bibitem{MD} H. M\"{uther} and W.H. Dickhoff, {\it Phys. Rev. C} {\bf 49}
(1994) R17; H. M\"{uther}, A. Polls, and W.H. Dickhoff, {\it Phys.
Rev. C} {\bf 51} (1995) 3051.

\bibitem{Amir} K. Amir-Azimi-Nili, H. M\"{u}ther, L.D. Skouras, and A. Polls,
{\it Nucl. Phys. A} {\bf 604} (1996) 245; K. Amir-Azimi-Nili, J.M.
Ud\'{\i}as, H. M\"{u}ther, L.D. Skouras, and A. Polls, {\it Nucl.
Phys. A} {\bf 625} (1997) 633.

\bibitem{Geurts} W.J.W. Geurts, K. Allaart, W.H. Dickhoff, and H. M\"{uther}, {\it Phys.
Rev. C} {\bf 53} (1996) 2207.

\bibitem{Schwandt}
P. Schwandt, H.O. Meyer, W.W. Jacobs, A.D. Bacher, S.E. Vigdor,
M.D. Kaitchuck, and T.R. Donoghue, {\it Phys. Rev. C} {\bf 26}
(1982) 55.

\bibitem{Kra90} J. Kramer, Ph.D. thesis, Universiteit van Amsterdam, 1990.

\bibitem{Abe92} C. Van den Abeele, D. Ryckbosch, J. Ryckebusch, J. Dias, L.
Van Hoorebeke, R. Van de Vyver, J.-O. Adler, B.-E. Andersson, L.
Isaksson, H. Ruijter, and B. Schr\"{o}der, {\it Phys. Lett. B}
{\bf 296} (1992) 302.

\bibitem{Gai2000} M.K. Gaidarov, K.A. Pavlova, A.N. Antonov, M.V.
Stoitsov, S.S. Dimitrova, M.V. Ivanov, and C. Giusti, {\it Phys.
Rev. C} {\bf 61} (2000) 014306.

\bibitem{Iva2001} M.V. Ivanov, M.K. Gaidarov, A.N. Antonov, and
C. Giusti, {\it Phys. Rev. C} {\bf 64} (2001) 014605.

\bibitem{Wes92} J. Wesseling, C.W. de Jager, L. Lapik\'{a}s, H. de
Vries, M.N. Harakeh, N. Kalantar-Nayestanaki, L.W. Fagg, R.A.
Lindgren, and D. Van Neck, {\it Nucl. Phys. A} {\bf A547} (1992)
519.

\end{thebibliography}
\end{document}